\documentclass[]{pasj02} 
\usepackage[switch,mathlines]{lineno} 
\usepackage{natbib} 
\usepackage{caption}
\usepackage{comment}

\jyear{2026}
\Received{}
\Accepted{}

\begin{document} 

\title{MOA-2020-BLG-108Lb: A Giant Planet Beyond the Snow Line of a Low-Mass Lens Near the Lower Boundary of the Mass-Ratio Desert}

\author{
 Yuki \textsc{K. Satoh},\altaffilmark{1}$^{,}$\altaffilmark{2}\altemailmark\orcid{0000-0002-1228-4122} \email{yukisatoh.work@gmail.com} \email{yukisato@kanto-gakuin.ac.jp} \email{sato.yuki@nit.ac.jp} 
 David \textsc{P. Bennett},\altaffilmark{3}$^{,}$\altaffilmark{4}\orcid{0000-0001-8043-8413}
 Takahiro \textsc{Sumi},\altaffilmark{5}\orcid{0000-0002-4035-5012}
 Ian \textsc{A. Bond},\altaffilmark{6}\orcid{0000-0002-8131-8891}
 Nicholas \textsc{J. Rattenbury},\altaffilmark{7}\orcid{0000-0001-5069-319X}
 Daisuke \textsc{Suzuki},\altaffilmark{5}\orcid{0000-0002-5843-9433}
 Naoki \textsc{Koshimoto},\altaffilmark{5}\orcid{0000-0003-2302-9562}
 Shota \textsc{Miyazaki},\altaffilmark{8}\orcid{0000-0001-9818-1513}
 Rintaro \textsc{Kirikawa},\altaffilmark{5}\orcid{0009-0005-3016-4211}
 Fumio \textsc{Abe},\altaffilmark{9}
 Aparna \textsc{Bhattacharya},\altaffilmark{3}$^{,}$\altaffilmark{4}
 Ryusei \textsc{Hamada},\altaffilmark{5}
 Stela \textsc{Ishitani~Silva},\altaffilmark{3}$^{,}$\altaffilmark{10}\orcid{0000-0003-2267-1246}
 Yuki \textsc{Hirao},\altaffilmark{11}\orcid{0000-0003-4776-8618}
 Yutaka \textsc{Matsubara},\altaffilmark{9}
 Yasushi \textsc{Muraki},\altaffilmark{9}\orcid{0000-0003-1978-2092}
 Tutumi \textsc{Nagai},\altaffilmark{5}
 Kansuke \textsc{Nunota},\altaffilmark{5}\orcid{0009-0005-3414-455X}
 Greg \textsc{Olmschenk},\altaffilmark{3}\orcid{0000-0001-8472-2219}
 Cl$\boldsymbol{\acute{\rm e}}$ment \textsc{Ranc},\altaffilmark{12}\orcid{0000-0003-2388-4534}
 Sean \textsc{K. Terry},\altaffilmark{3}$^{,}$\altaffilmark{4}\orcid{0000-0002-5029-3257}
 Paul \textsc{J. Tristram},\altaffilmark{13}
 Aikaterini \textsc{Vandorou},\altaffilmark{3}$^{,}$\altaffilmark{4}\orcid{0000-0002-9881-4760}
  and 
 Hibiki \textsc{Yama}\altaffilmark{14}\orcid{0000-0001-7692-0581}
}
\altaffiltext{1}{College of Science and Engineering, Kanto Gakuin University, Yokohama, Kanagawa 236-8501, Japan}
\altaffiltext{2}{Department of Electrical and Information Engineering, Nippon Institute of Technology, Miyashiro, Saitama 345-8501, Japan}
\altaffiltext{3}{Code 667, NASA Goddard Space Flight Center, Greenbelt, MD 20771, USA}
\altaffiltext{4}{Department of Astronomy, University of Maryland, College Park, MD 20742, USA}
\altaffiltext{5}{Department of Earth and Space Science, The University of Osaka, Toyonaka, Osaka 560-0043, Japan}
\altaffiltext{6}{Institute of Natural and Mathematical Sciences, Massey University, Auckland 0745, New Zealand}
\altaffiltext{7}{Department of Physics, University of Auckland, Private Bag 92019, Auckland, New Zealand}
\altaffiltext{8}{Institute of Space and Astronautical Science, Japan Aerospace Exploration Agency, Sagamihara, Kanagawa 252-5210, Japan}
\altaffiltext{9}{Institute for Space-Earth Environmental Research, Nagoya University, Nagoya, Aichi 464-8601, Japan}
\altaffiltext{10}{Department of Physics, The Catholic University of America, Washington, DC 20064, USA}
\altaffiltext{11}{Institute of Astronomy, The University of Tokyo, Mitaka, Tokyo 181-0015, Japan}
\altaffiltext{12}{Sorbonne Universit$\acute{\rm e}$, CNRS, UMR 7095, Institut d’Astrophysique de Paris, 98 bis bd Arago, F-75014 Paris, France}
\altaffiltext{13}{University of Canterbury Mt. John Observatory, P.O. Box 56, Lake Tekapo 8770, New Zealand}
\altaffiltext{14}{Department of Astronomy, Kyoto University, Kyoto, Kyoto 606-8502, Japan}




\maketitle

\begin{abstract}
\label{Abstract}
We present an analysis of the microlensing event MOA-2020-BLG-108, which was discovered in June 2020 by the MOA collaboration toward the Galactic bulge.
The observed light curve shows significant deviations from the standard single-lens single-source model.
We find two degenerate binary-lens single-source solutions, corresponding to the wide and close configurations, with a companion-to-host mass ratio of $q\sim0.02$ and projected host--companion separations of $s=1.33\pm0.01$ and $s=0.76\pm0.01$, respectively.
These solutions improve the fit by $\Delta\chi^2>4430$ compared to the single-lens model.
We detected the finite-source effect in the light curve and obtained the angular Einstein radius of $\theta_{\rm E} = 0.7\pm0.1\:\mathrm {mas}$, which provides a mass--distance relation for the lens.
We conducted a Bayesian analysis to estimate the physical parameters of the lens system.
The results indicate that the lens system consists of a host star with a mass of $M_{\rm L,H} \sim 0.6\:M_\odot$ at a distance of $D_{\rm L}\sim5$ kpc and a giant planet with a mass of $M_{\rm L,C}\sim10\:M_{\rm {Jup}}$
orbiting beyond the snow line.
Conventional planet formation theories suggest that giant planets are unlikely to form around low-mass stars.
Furthermore, several statistical studies have suggested the existence of a companion-to-host mass-ratio desert in the range $0.02 \lesssim q \lesssim 0.05$, and the companion in the lens system discovered in this work lies near the lower boundary of this desert.
Objects near the planet-brown dwarf boundary may form through multiple pathways, and this discovery provides an additional data point for understanding their formation mechanisms.
\end{abstract}


\section{Introduction}
\label{sec:Introduction}
Gravitational microlensing is a method for detecting planets that utilizes the gravitational bending of light, as predicted by the general theory of relativity \citep{Liebes1964, Paczynski1991}.
In gravitational microlensing, the brightness of a background source star varies over time as a foreground gravitating object (hereafter referred to as the lens) passes close to the line of sight to the source.
For a single-lens system, the source star reaches its maximum brightness when the observer, the lens, and the source star are collinear.
In a two-component lens system, the gravity of the companion perturbs the magnification produced by the host star.
A companion such as a planet can be detected by observing such perturbations in the light curve \citep{Mao+1991}.
The event rate of gravitational microlensing is approximately $10^{-5}$  events $\rm{yr}^{-1} \rm{star}^{-1}$, and for planetary microlensing events, it is further suppressed by a factor of ~ $10^{-2}$.
Therefore, most gravitational microlensing observations for planet searches target the Galactic bulge \citep[e.g.,][]{Udalski2003, Bond+2001, Kim+2016}, which enables efficient surveys by allowing simultaneous monitoring of a large number of stars.
Perturbations induced by a companion in the lens system are most prominent when the companion lies near the Einstein radius of the host star.
Since the Einstein radii of typical lens stars toward the Galactic bulge are $\sim 2 - 6$ au, microlensing is sensitive to planets around or beyond the snow line that are difficult for other methods, such as radial-velocity or transit methods.
Planet formation theories suggest that the region around the snow line, where solid materials are abundant, is an efficient site for the formation of giant and ice planets \citep[e.g.,][]{Ida+2004, Laughlin+2004, Kennedy+2006}.
Therefore, this region is crucial for understanding the global picture of planet formation.

Statistical analyses of gravitational microlensing events \citep{Shvartzvald+2016, Suzuki+2016, Zang+2025} indicate that, in the planetary-mass regime, the lens companion-to-host mass-ratio $q$ distribution declines logarithmically toward larger $q$.
\noindent \citet{Shvartzvald+2016} examined the distribution of mass ratios, $q$, and identified a pronounced deficit around $q \sim 10^{-2}$, which appears to separate the populations of stellar binaries and planetary companions.
Here the occurrence rate of Jupiter-mass planets is typically about an order of magnitude lower than that of Neptune-mass planets, with a minimum near the super-Jupiter mass.
Using 63 planets from 60 gravitational microlensing events observed by the Korea Microlensing Telescope Network \citep[KMTNet;][]{Kim+2016} between 2016 and 2019, \citet{Zang+2025} derived the planetary mass-ratio function over $-5.2 \leq \log q \leq -1.5$ and demonstrated that the planet frequency declines toward higher mass ratios.
In the double-Gaussian model, motivated by the distinct formation pathways of rocky and gas planets, planets with mass ratios greater than $10^{-2}$ occur at a rate less than one-tenth of that for planets with mass ratios around $10^{-2.5}$.
Furthermore, \citet{Zhang2025} identified a mass-ratio desert at $0.02 \lesssim q \lesssim 0.05$ for projected separations of 1--5 au based on a statistical analysis of binary-lens microlensing events. 
They also found a statistically significant truncation in the mass-ratio distribution at $q \sim 0.02$, above which the occurrence rate declines by about an order of magnitude.
Outside of microlensing studies, the existence of a mass-ratio desert at $0.02 \lesssim q \lesssim 0.05$ has also been demonstrated in direct-imaging statistical surveys of intermediate-mass ($1.75-4.5\:M_\odot$) stars at orbital separations of $20-1000$~au \citep{Duchene+2023}.
More recently, \citet{Giacalone+2026} performed a joint analysis of radial-velocity and astrometric data for companions at separations of $1-50$~au and found a transition in the mass-ratio distribution at $q = 0.027_{-0.010}^{+0.015}$, suggesting a possible transition in formation pathways.

We present an analysis of the gravitational microlensing event MOA-2020-BLG-108, from which we identify a giant planet orbiting the lens star.
Section~\ref{sec:Observation_and_Data_Reduction} describes the observations and the data reduction procedures.
Section~\ref{sec:Light-curve_Modeling} presents the details of the light curve modeling.
Section~\ref{sec:Source_Color_and_Magnitude} describes the color and magnitude of the source star and the angular Einstein radius.
Section~\ref{sec:Lens_System_Properties}  presents an estimation of the properties of the lens system based on a Bayesian analysis.
Section~\ref{sec:Discussion_and_Conclusion} provides the discussion and conclusions. 

\section{Observation and Data Reduction}
\label{sec:Observation_and_Data_Reduction}
\begin{figure*}
    \centering
    \hspace*{-18mm}
    \includegraphics[scale=0.65,angle=-90]{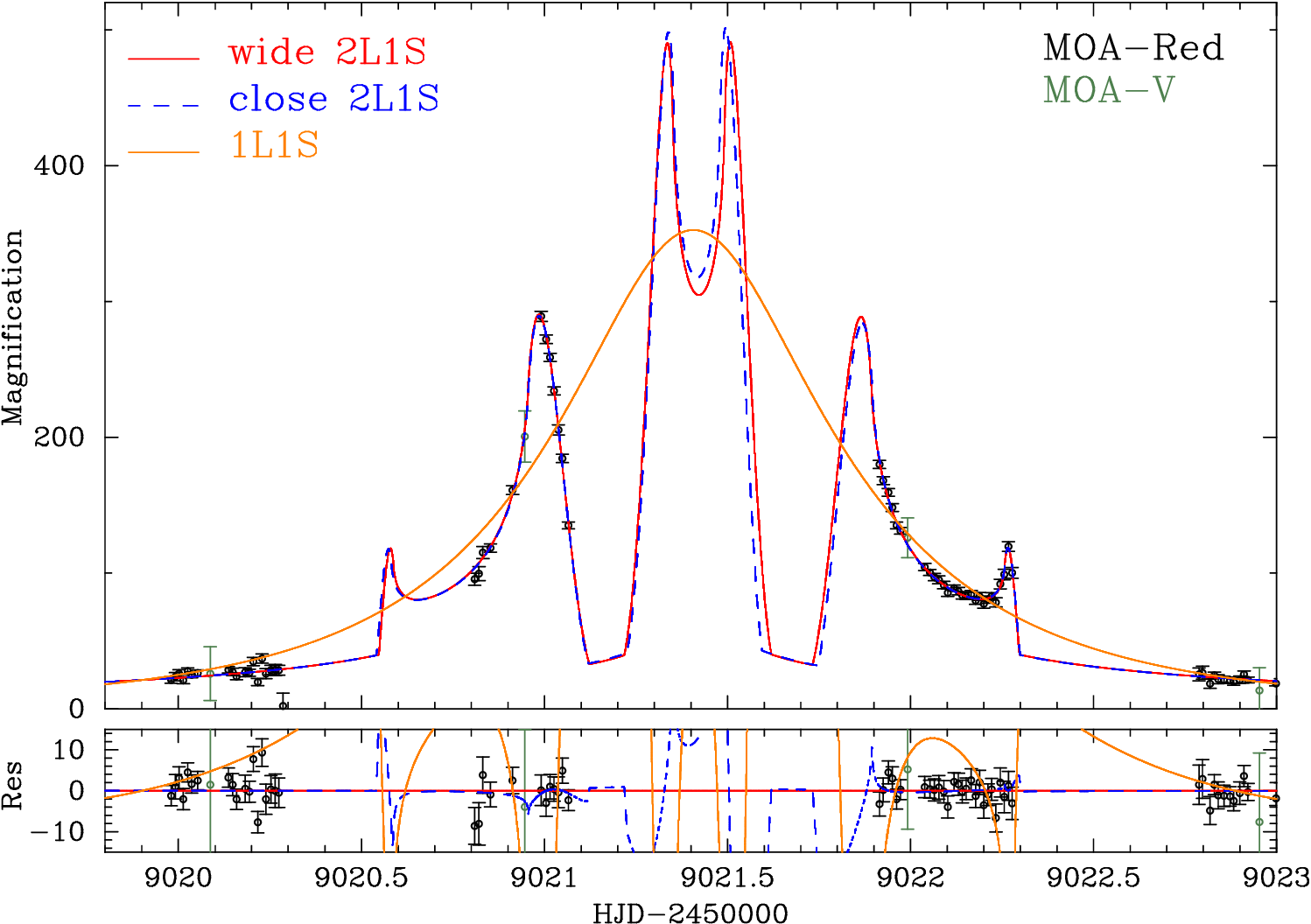}
    \vspace{10mm}
    \hspace{-1mm}
    \caption{(Top panel) Light curve of MOA-2020-BLG-108.
    The error bars are renormalized according to Equation~(\ref{eq:error}).
    The red solid, blue dashed, and yellow solid lines represent the wide 2L1S, close 2L1S, and 1L1S models, respectively.
    (Bottom panel) Residuals from the wide 2L1S model. \\
    {Alt text: Two vertically stacked panels with time on the horizontal axes. The bottom panel shows the differences obtained by subtracting the wide 2L1S model curve from the light curve, the 1L1S model curve, and the close 2L1S model curve.}
    }
    \label{fig:lightcurve}
\end{figure*}

The microlensing event MOA-2020-BLG-108 was discovered by the Microlensing Observations in Astrophysics \citep[MOA;][]{Bond+2001,Sumi+2003} Collaboration on 2020 June 20 (${\rm HJD}^{\prime}\sim9021$)\footnote{${\rm HJD}^{\prime}\equiv{\rm HJD}-2450000$} through the MOA alert system \citep{Bond+2001}.
Since May 2005, the MOA Collaboration has conducted a long-term microlensing survey toward the Galactic bulge using the $1.8$-m MOA-$\rm{I\hspace{-.01em}I}$ telescope at the University of Canterbury’s Mount John Observatory in New Zealand.
The MOA-$\rm{I\hspace{-.01em}I}$ telescope is equipped with the MOA-Cam3 camera \citep{Sako+2008}, at its prime focus, which consists of ten $2\rm{k} \times 4\rm{k}$ pixel CCD (Charge-Coupled Device) image sensors, providing a wide field of view of $2.2$ square degrees.
This enables the MOA-$\rm{I\hspace{-.01em}I}$ telescope to monitor an area of approximately $50$ square degrees toward the Galactic bulge at a high cadence of once every $15-90$ minutes.
This event occurred at J2000 equatorial coordinates $(\alpha,\delta)_{\mathrm{J2000}}=(17^h 58^m 51^s.74, -26^\circ 54^\prime 59^{\prime\prime}.43)$, corresponding to Galactic coordinates $(l, b) = (3.228, -1.488)$.
This position is located in the MOA-$\rm{I\hspace{-.01em}I}$ field gb10, which is observed with a $15$-minute cadence.
Observations with the MOA-$\rm{I\hspace{-.01em}I}$ telescope are primarily conducted in the $R_{\rm {MOA}}$ band, which corresponds to a wavelength range combining the standard Kron–Cousins $R$ and $I$ bands, while observations in the Johnson $V_{\rm {MOA}}$ band are performed approximately once per day for the purpose of measuring color information.

The light-curve data are reduced using a photometric pipeline based on the difference image analysis  \citep{Tomaney+1996, Alard+1998}, as implemented by \citet{Bond+2001}.
This pipeline typically underestimates the photometric uncertainties.
To properly assess the uncertainties, we apply the standard empirical error normalization procedure described by \citet{Bennett+2008} and \citet{Yee+2012}, as given by the following equation:

\begin{equation}\label{eq:error}
    \sigma^{\prime}_i = k\sqrt{\sigma^2_i + e^2_{\rm min}},
\end{equation}

\noindent where $\sigma^{\prime}_i$ denotes the renormalized uncertainty and $\sigma_i$ is the original uncertainty provided by the photometric pipeline.
The parameters $e_{\mathrm{min}}$ and $k$ are variables used for the error renormalization.
First, a preliminary best-fit model is obtained from an initial analysis using $\sigma_i$.
The parameter $e_{\mathrm{min}}$ is then chosen so as to linearize the slope of the cumulative $\chi^2$ as a function of magnification derived from this model, and $k$ is selected such that the reduced $\chi^2 \simeq 1$. 
This procedure is iterated each time an improved best-fit model is found.
This procedure accounts for low-level unknown systematic errors and does not affect the parameter estimation in the light-curve modeling \citep{Ranc+2019}.
In this analysis, we used data in the range $8140 \le \mathrm{HJD}^{\prime} \le 9160$, corresponding to three years including the magnification peak, in order to remove long-term systematic trends from these data sets.
Table~\ref{tab:dataset} shows the data sets and the renormalization parameters.
Figure~\ref{fig:lightcurve} shows the light curve of MOA-2020-BLG-108, along with the single-lens single-source model (hereafter 1L1S) and the binary-lens single-source model (hereafter 2L1S).
For the 2L1S model, we plot the best-fit wide 2L1S solution and its degenerate close 2L1S solution.

\begin{table}
  \tbl{Data Sets for MOA-2020-BLG-108}{
  \begin{tabular}{llccc}
      \hline
      Filter & Label & $N_{\rm {use}}$ & $k$\footnotemark[$a$] & $e_{\rm {min}}$\footnotemark[$a$]  \\ 
      \hline
      $R_{\rm {MOA}}$ & MOA-Red & 5897 & 1.151 & 0.005 \\
      $V_{\rm {MOA}}$ & MOA-V & 208 & 1.151 & 0.003 \\
      \hline
    \end{tabular}}\label{tab:dataset}
\begin{tabnote}
\footnotemark[$a$] Parameters for the error normalization.  \\ 
\end{tabnote}
\end{table}

\section{Light-curve Modeling}
\label{sec:Light-curve_Modeling}
\subsection{Single Lens Model Analysis}
\label{subsec:Single_Lens_Model_Analysis}

The model flux for a microlensing event is given by the following equation:

\begin{equation}\label{eq:magnification}
    f(t)=A(t,\boldsymbol{x})f_{\rm s} + f_{\rm b},
\end{equation}

\noindent where $t$ is time, $f(t)$ is the total flux, $f_{\rm s}$ is the source flux, $f_{\rm b}$ is the blended flux, and $A(t,\boldsymbol{x})$ is the magnification.
In the analysis of the 1L1S model, $\boldsymbol{x}$ consists of four parameters \citep{Paczynski1986}: the time of closest approach to the center of mass, $t_0$; the Einstein radius crossing time, $t_{\rm{E}}$; the impact parameter, $u_0$, and the normalized angular source radius, $\rho$.
Both $u_0$ and $\rho$ are in units of the angular Einstein radius, $\theta_{\rm E}$.
The parameters $f_{\rm s}$ and $f_{\rm b}$ are obtained using the linear fitting method described by \citet{Rhie+1999}.
We explore the parameter space using the Metropolis--Hastings algorithm.
Finite-source effects, arising from the finite angular radius of the source star, are computed using the image-centered inverse ray-shooting method \citep{Bennett+1996, Bennett+2010b} implemented by \citet{Sumi+2010}.

We adopt the following expression for the limb-darkening law of the source star:

\begin{equation}\label{eq:limb-darkening}
    S_\lambda(\vartheta) = S_\lambda(0)\left[1-u_\lambda(1-\cos(\vartheta))\right],
\end{equation}

\noindent where $\vartheta$ denotes the angle between the line of sight and the normal to the surface of the source star, and $S_\lambda(\vartheta)$ is the limb-darkened surface brightness at angle $\vartheta$ and observed wavelength $\lambda$. From the source color and magnitude described in Section~\ref{sec:Source_Color_and_Magnitude}, we estimate an effective temperature of $T_{\rm eff} = 4930 \pm 493\,\mathrm{K}$ \citep{Gonzalez+2009}.
We assume a source metallicity of $[\mathrm{M/H}]=0$, a surface gravity of $\log g = 4.5$, and a microturbulent velocity of $v = 1\,\mathrm{km\,s^{-1}}$, and adopt limb-darkening coefficients of $u_V = 0.7442$, $u_R = 0.6585$, and $u_I = 0.5624$ based on the ATLAS models of \citet{Claret+2011}.
As described in Section~\ref{sec:Observation_and_Data_Reduction}, the \(R_{\mathrm{MOA}}\) band of the MOA-$\rm{I\hspace{-.01em}I}$ telescope spans a wavelength range combining the Kron--Cousins \(R\) and \(I\) bands, and thus we adopt
\(u_{R_{\mathrm{MOA}}} = (u_R + u_I)/2 = 0.61045\).
We performed a light-curve analysis using the 1L1S model.
However, this solution is worse than the best-fit solution of the 2L1S model described in Section~\ref{subsec:Binary_Lens_Model_Analysis} by $\Delta\chi^2 = 4436$.
The best-fit parameters of these models are summarized in Table~\ref{tab:fit_param}.

We also tested several single-lens-based models, including models with higher-order effects and a binary-source model.
First, we examined a model including microlens parallax, which accounts for the effect of the Earth's orbital motion on the light curve \citep{Gould1992, Gould2004, Smith+2003, Dong+2009}.
This model requires two additional parameters, the microlens parallax vector $\boldsymbol{\pi}_{\mathrm E} =
(\pi_{{\mathrm E},{\mathrm N}}, \pi_{{\mathrm E},{\mathrm E}})$ \citep{Gould2000}.
We then examined a 1L1S model including the xallarap effect.
The xallarap effect accounts for the orbital motion of the source star and is parameterized by the orientation of the source orbit, specified by $\mathrm{R.A.}_{\xi}$ and $\mathrm{decl.}_{\xi}$; the source orbital period, $P_{\xi}$; the source orbital eccentricity, $e_{\xi}$; the time of periastron passage, $T_{\mathrm{peri},\xi}$; and the xallarap vector, $\boldsymbol{\xi}_{\mathrm{E}} = (\xi_{\mathrm{E},\mathrm{N}}, \xi_{\mathrm{E},\mathrm{E}})$.
We also tested a single-lens binary-source model (hereafter 1L2S), in which two source stars are lensed by a single lens and the observed light curve is represented by the sum of their magnified fluxes.
In addition to the standard 1L1S parameters, this model introduces parameters for the second source: its time of closest approach to the lens, $t_{0,2}$; impact parameter, $u_{0,2}$; normalized angular source radius, $\rho_2$; and the flux ratio, $q_{\mathrm{F},j}$, in each passband $j$.
Finally, we tested a 1L2S model including microlens parallax.
Although these single-lens-based models improve the fit relative to the standard 1L1S model, they still provide substantially worse fits than the 2L1S model in Section~\ref{subsec:Binary_Lens_Model_Analysis}.

\subsection{Binary Lens Model Analysis}
\label{subsec:Binary_Lens_Model_Analysis}
\begin{table*}
  \tbl{Best-Fit Parameters and 1 $\sigma$ Uncertainties for the 1L1S and 2L1S Models}{
  \begin{tabular}{ccccc}
      \hline
      Parameter & Units & 1L1S & wide 2L1S & close 2L1S  \\ 
      \hline
      $t_0$ & $\rm {HJD}-24,59,020$ & $1.407\pm0.002$ & $1.422\pm0.007$ & $1.417\pm0.007$ \\
      $t_{\rm {E}}$ & days & $0.86\pm0.04$ & $31.56\pm5.27$ & $34.02\pm4.06$ \\
      $u_{\rm {0}}$ & $10^{-3}$ & $442.48\pm43.55$ & $5.29\pm0.82$ & $4.90\pm0.65$ \\
      $q$ & $10^{-2}$ & - & $1.74\pm0.25$ & $1.57\pm0.19$ \\
      $s$ &  & - & $1.33\pm0.01$ & $0.76\pm0.01$ \\
      $\alpha$ & radians & - & $4.71\pm0.01$ & $4.71\pm0.01$ \\
      $\rho$ & $10^{-4}$ & $325.21\pm150.66$ & $6.12\pm1.03$ & $5.73\pm0.85$ \\
      \hline
      $\chi^2$ &  & $10535.7$ & $6100.2$ & $6102.1$ \\
      $\Delta \chi^2 $ &  & $4435.6$ & - & $1.9$ \\
      \hline
    \end{tabular}}\label{tab:fit_param}
\end{table*}

Describing the magnification of the standard 2L1S model requires three additional parameters: the mass ratio of a lens companion relative to the host, $q$; the projected separation between the host star and the companion normalized to the Einstein radius, $s$; and the angle between the binary-lens axis and the direction of the source trajectory, $\alpha$.
To identify the global minimum in the highly complex, high-dimensional $\chi^2$ parameter space, we first carried out a grid search over 34,440 combinations of the parameters $(q,\: s,\: \alpha)$, which most strongly affect the light-curve morphology.
Here we uniformly take $21$ values in the range $-5 \leq \log q \leq 0$, $41$ values in the range $-1.25 \leq \log s \leq 1.25$, and $40$ values in the range $0 \leq \alpha \leq 2 \pi$.
During this process, all other parameters were allowed to vary freely.

\begin{figure*}
    \centering
    \includegraphics[scale=0.45,angle=0]{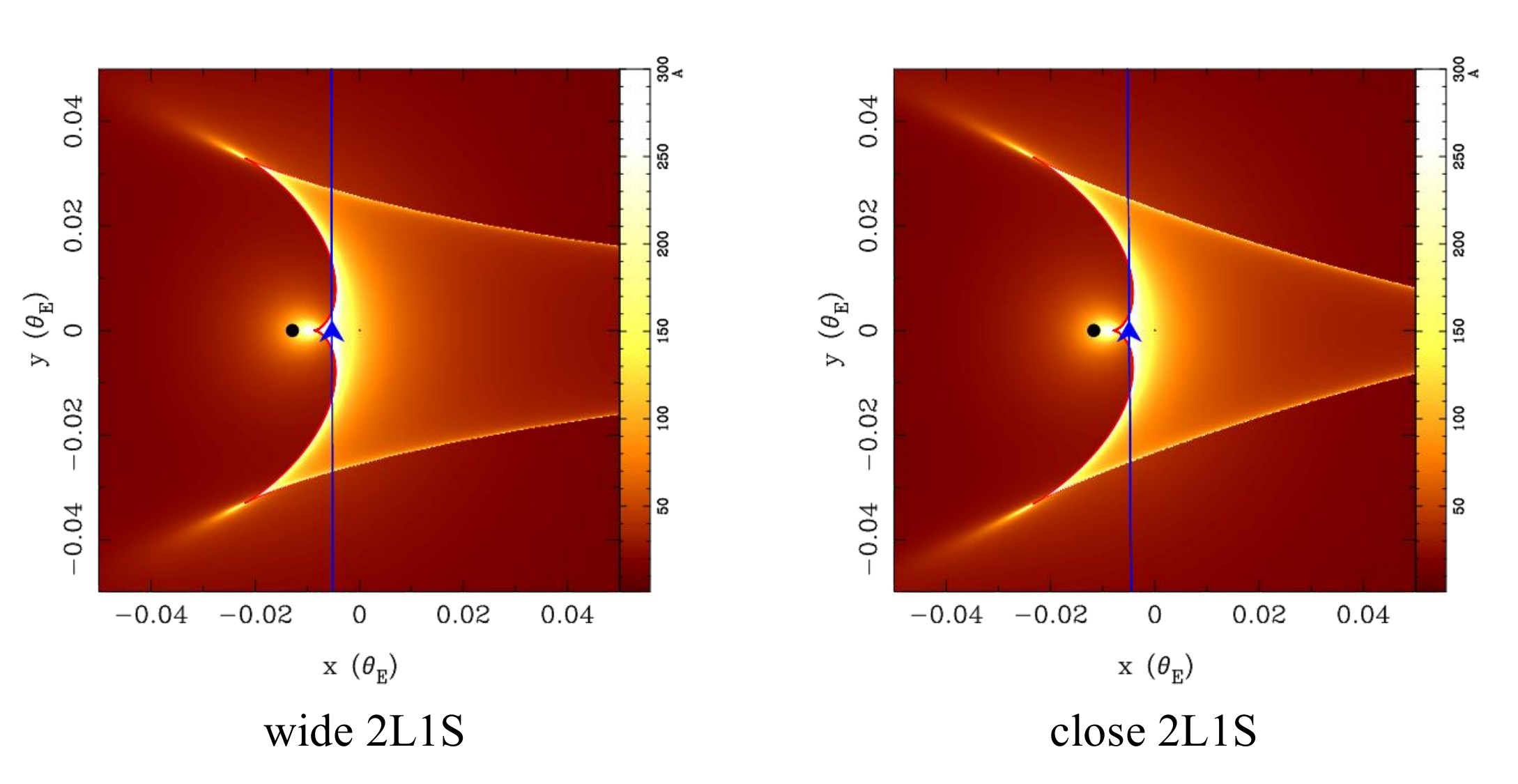}
    \caption{The position of the lens host, the source trajectory, and caustics on the magnification maps for the wide 2L1S solution (left panel) and the close 2L1S solution (right panel).
    The black filled circle indicates the lens host.
    The blue line with an arrow represents the source trajectory.
    The red curve represents the caustic.
    The colored contours represent the magnification map. \\
    {Alt text: Two side-by-side panels showing the geometry for the wide and close 2L1S solutions.}
    }
    \label{fig:caustic}
\end{figure*}

Subsequently, a detailed fitting was performed by allowing all parameters to vary freely for all models within $\Delta\chi^2 = 2000$ of the best-fit solution obtained from the initial grid search.
As a result, we found that two 2L1S models with $(q, s) = (1.74 \times 10^{-2}, 1.33)$ (wide model) and $(q, s) = (1.57 \times 10^{-2}, 0.76)$ (close model) reproduce the light curve well.
As summarized in Table~\ref{tab:fit_param}, the wide 2L1S solution has $\chi^2 = 6100.2$, while the close 2L1S solution has $\chi^2 = 6102.1$, corresponding to $\Delta\chi^2 = 1.9$.
Figure~\ref{fig:caustic} shows the position of the lens host, source trajectories, and caustic structures on the magnification maps for the wide and close 2L1S solutions.

We further tested models that include the higher-order effects of parallax and xallarap, as described in Section~\ref{subsec:Single_Lens_Model_Analysis}, together with lens orbital motion.
The lens orbital-motion effect can be modeled by introducing the time derivatives of the projected separation between the lens components, $ds/dt$, and of the angle of the source trajectory, $d\alpha/dt$, both evaluated at $t_0$.
For the lens orbital-motion model, we imposed the constraint that the ratio of kinetic to potential energy is less than unity.
We searched for best-fit models including these higher-order effects and compared them using both $\chi^2$ and the Bayesian Information Criterion (BIC).
In Table~\ref{tab:Model_Comparison}, we list $\Delta\chi^2$ and $\Delta\mathrm{BIC}$, which represent the differences relative to the best wide 2L1S solution.
We found that the improvements obtained by including these higher-order effects are not significant.
Thus, we adopted the wide and close 2L1S solutions as the best models and use them in the subsequent analyses.

\begin{table}
  \tbl{$\Delta \chi^2$ and $ \Delta \rm{BIC}$ for Each Model Relative to the wide 2L1S Solution}{
  \begin{tabular}{lccc}
      \hline
      Model & $N_{\mathrm{param}}$\footnotemark[$a$] & $\Delta\chi^2$ & $\Delta\mathrm{BIC}$   \\ 
      \hline
      1L1S & $8$ & $4435.6$ & $4409.4$  \\
      1L1S+parallax & $10$ & $4152.5$ & $4143.8$  \\
      1L1S+xallarap  & $15$ & $577.6$ & $612.4$  \\
      1L2S & $13$ & $4239.0$ & $4256.4$  \\
      1L2S+parallax  & $15$ & $3751.9$ & $3786.7$  \\
      2L1S (wide) & $11$ & - & -  \\
      2L1S+parallax  & $13$ & $-3.4$ & $14.0$  \\
      2L1S+xallarap  & $18$ & $-11.9$ & $49.1$  \\
      2L1S+parallax+LOM  & $15$ & $-14.8$ & $20.1$  \\
      \hline
    \end{tabular}}\label{tab:Model_Comparison}
    \begin{tabnote}
    \footnotemark[$a$] The number of free parameters used in each model.
    \par
    ``+parallax'', ``+xallarap'', and ``+LOM'' denote models including the microlens parallax effect, the xallarap effect, and lens orbital motion, respectively.
    \end{tabnote}
\end{table}

\section{Source Color and Magnitude}
\label{sec:Source_Color_and_Magnitude}
To estimate the physical parameters of the lens through Bayesian analysis, we first determine the properties of the source star and then derive the angular Einstein radius, $\theta_{\rm E}$.
We calibrate the instrumental magnitudes in the MOA $R_{\mathrm{MOA}}$ and $V_{\mathrm{MOA}}$ bands to the standard Cousins $I$ and Johnson $V$ systems by cross-matching MOA catalog stars within $0.4$ arcsec of the event position with the OGLE-$\rm{I\hspace{-.15em}I\hspace{-.15em}I}$ photometric map \citep{Szymanski+2011}.
For reliability, we restricted stars to $16 \leq V_{\mathrm{OGLE-\rm{I\hspace{-.15em}I\hspace{-.15em}I}}} \: [\mathrm{mag}] \leq 19$, and performed $5\:\sigma$ clipping in the linear regressions.
From the final $70$ remaining objects, the following conversion equations from $R_{\mathrm {MOA}}$ and ${(V-R)}_{\mathrm {MOA}}$ to $I_{\mathrm{OGLE-\rm{I\hspace{-.15em}I\hspace{-.15em}I}}}$ and ${(V - I)}_{\mathrm{OGLE-\rm{I\hspace{-.15em}I\hspace{-.15em}I}}}$ were obtained by linear regression:

\begin{equation}\label{eq:convert1}
\begin{aligned}
I_{\rm OGLE-\mathrm{I}\hspace{-1.2pt}\mathrm{I}\hspace{-1.2pt}\mathrm{I}}
&= R_{\rm MOA}
   -(0.190\pm0.003)\times (V-R)_{\rm MOA} \\
&\quad + (28.289\pm0.004),
\end{aligned}
\end{equation}

\begin{equation}\label{eq:convert2}
\begin{aligned}
(V-I)_{\rm OGLE-\mathrm{I}\hspace{-1.2pt}\mathrm{I}\hspace{-1.2pt}\mathrm{I}}
&= (1.111\pm0.006)\times (V-R)_{\rm MOA} \\
&\quad + (0.198\pm0.009).
\end{aligned}
\end{equation}

Using the above transformation relations, we obtain the apparent color and magnitude of the source star, $(V-I, I)_{\mathrm S}=(3.186\pm0.090,22.582\pm0.018)$ for the wide 2L1S solution and $(V-I, I)_{\mathrm S}=(3.175\pm0.091,22.676\pm0.018)$ for the close 2L1S solution.
We estimate the centroid of the red clump giant (RCG) population in the $(V - I, I)$ color–magnitude diagram (CMD) within $2$~arcmin of the source star to be $(V - I, I)_{\mathrm {RCG}} = (3.266 \pm 0.016, 16.954 \pm 0.088)$.
The intrinsic RCG centroid in this direction is estimated to be $(V - I,I)_{\mathrm {RCG,0}} = (1.060 \pm 0.060, 14.443 \pm 0.040)$\citep{Bensby+2013, Nataf+2013b}.
From this, we derive the reddening and extinction toward the event as $(E(V - I), A(I)) = (2.206 \pm 0.062, 2.511 \pm 0.097)$.

\begin{figure}
    \centering
    \includegraphics[scale=0.37]{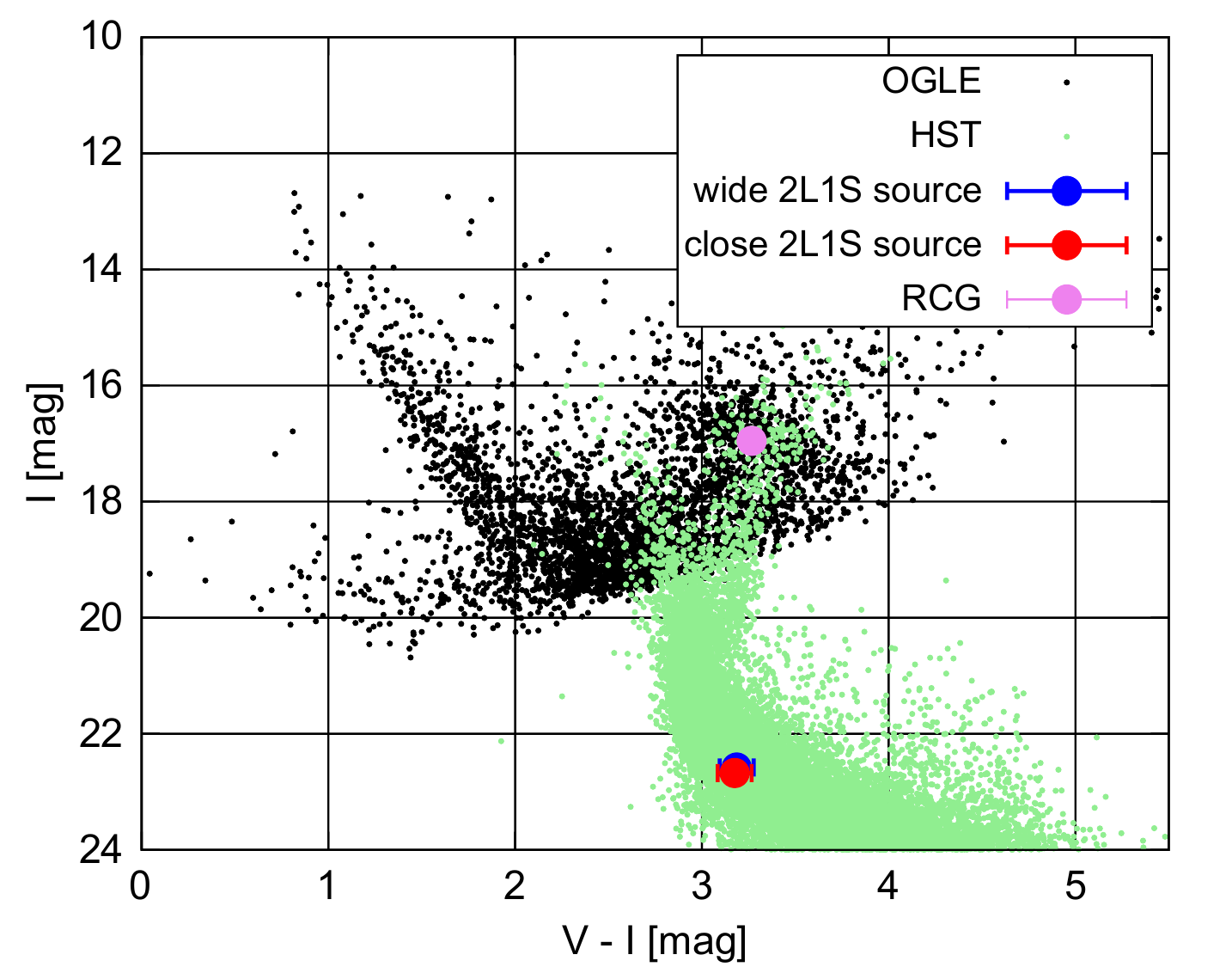}
    \caption{
    Color-Magnitude Diagram (CMD, black dots) of the OGLE-$\mathrm{I}\hspace{-1.2pt}\mathrm{I}\hspace{-1.2pt}\mathrm{I}$ stars within $2$~arcmin of MOA-2020-BLG-108.
    The green dots denote stars in Baade's Window observed with the Hubble Space Telescope \citep{Holtzman+1998}, transformed to the CMD by matching the color and magnitude of the RCG centroid. 
    The blue and red circles represent the source positions for the wide and close 2L1S solutions, respectively, while the pink circle represents the centroid of the RCGs within $2$~arcmin of MOA-2020-BLG-108.\\
    {Alt text: A color magnitude diagram with V minus I color on the horizontal axis and I band magnitude on the vertical axis.}
    }
    \label{fig:cmd}
\end{figure}

We finally derive that the extinction-corrected apparent color and magnitude of the source star are $(V-I, I)_{\mathrm{S},0} = (0.980 \pm 0.110,\, 20.072 \pm 0.099)$ for the wide 2L1S solution and $(V-I,\, I)_{\mathrm{S},0} = (0.969 \pm 0.110,\, 20.165 \pm 0.099)$ for the close 2L1S solution.
Figure~\ref{fig:cmd} shows the apparent CMD of stars in the OGLE-$\rm{I\hspace{-.15em}I\hspace{-.15em}I}$ photometric catalog within $2$~arcmin of the event, along with RCG centroid and the best-fit source position from the light curve modeling.
The green dots denote stars in Baade’s Window observed with the Hubble Space Telescope \citep{Holtzman+1998}, transformed to the CMD by matching the color and magnitude of the RCG centroid.
The source star of this event lies within the distribution of main-sequence stars in the HST data.
Using Table~\ref{tab:Lens_Properties} of \citet{Gonzalez+2009} with the source-star properties assumed in Section~\ref{subsec:Single_Lens_Model_Analysis}, we estimate an effective temperature of $T_{\rm eff} = 4930 \pm 493~\mathrm{K}$ from the value of $(V-I, I)_{\mathrm{S},0}$.
This result indicates that the source star is a late G-type or early K-type main-sequence star.
Finally, we estimate the angular source radius, $\theta_*$, using the following relation:

\begin{equation}\label{eq:theta_star}
    \log(2\theta_*)=0.50 + 0.42(V-I)_0 - 0.2 I_0,
\end{equation}

\noindent where this relation has an accuracy better than 2\% \citep{Fukui+2015}.
This relation is based on \citet{Boyajian+2014}, but was derived by limiting to FGK stars with $3900$ $<T_{\rm {eff}}$ [K] $<7000$ (Boyajian, 2014, Private Communication).
We then calculate the Einstein radius of the lens, $\theta_{\mathrm E} = \theta_* / \rho$, and the relative proper motion between the lens and the source, $\mu_{\mathrm {rel}} = \theta_{\mathrm E}/t_{\mathrm E}$.
As a result, we obtained $(\theta_*, \theta_{\mathrm E}, \mu_{\mathrm {rel}}) = (0.40\pm0.05\:\mathrm {\mu as}, 0.65\pm0.13\:\mathrm {mas},7.5\pm2.0\:\mathrm {mas \: yr^{-1}})$ for the wide 2L1S solution and $(\theta_*, \theta_{\mathrm E}, \mu_{\mathrm {rel}}) = (0.38\pm0.04\:\mathrm {\mu as}, 0.65\pm0.12\:\mathrm {mas},7.0\pm1.6\:\mathrm {mas \: yr^{-1}})$ for the close 2L1S solution.
The properties of the source star are summarized in Table~\ref{tab:Source_Properties}.

\begin{table}
  \tbl{Source Properties of the 2L1S Models}{
  \begin{tabular}{lccc}
      \hline
      Parameters & Units & wide 2L1S & close 2L1S   \\[2pt]
      \hline
      $I_{\mathrm S}$ & $\mathrm {mag.}$ & $22.58\pm0.02$ & $22.68\pm0.02$  \\[2pt]
      ${(V-I)}_{\mathrm S}$ & $\mathrm {mag.}$ & $3.19\pm0.09$ & $3.18\pm0.09$  \\[2pt]
      $I_{\mathrm S,0}$ & $\mathrm {mag.}$ & $20.07\pm0.10$ & $20.17\pm0.10$  \\[2pt]
      ${(V-I)}_{\mathrm S,0}$ & $\mathrm {mag.}$ & $0.98\pm0.11$ & $0.97\pm0.11$  \\[2pt]
      $\theta_*$ & $\mathrm {\mu as}$ & $0.40\pm0.05$ & $0.38\pm0.04$  \\
      $\theta_{\mathrm E}$ & $\mathrm {mas}$ & $0.65\pm0.13$ & $0.65\pm0.12$  \\[2pt]
      $\mu_{\mathrm{rel}}$ & $\mathrm {mas \: yr^{-1}}$ & $7.49\pm1.98$ & $7.02\pm1.57$  \\[2pt]
      \hline
    \end{tabular}}\label{tab:Source_Properties}
\end{table}

\section{Lens System Properties}
\label{sec:Lens_System_Properties}

The distance from Earth to the lens system, $D_{\mathrm L}$, and the total mass of the lens system, $M_{\mathrm L}$, are related through the following relations \citep{Gould2000}:

\begin{equation}
\label{eq:D_L}
    D_{\rm L}=\frac{\rm au}{\pi_{\rm E}\theta_{\rm E}+\pi_{\rm S}},
\end{equation}

\begin{equation}\label{eq:M_L}
    M_{\rm L}=\frac{\theta_{\rm E}}{\kappa \pi_{\rm E}},
\end{equation}

\noindent
where $\kappa = 4G/(c^2\,\mathrm{au}) \simeq 8.144~\mathrm{mas}~M_\odot^{-1}$ is a constant, and $\pi_{\mathrm S}$ is the parallax of the source star, given by $\pi_{\mathrm S} = \mathrm{au}/D_{\mathrm S}$.

As the parallax effect is not significantly detected for this event, we perform a Bayesian analysis \citep{Beaulieu+2006, Gould+2006, Bennett+2008} to estimate the physical parameters of the lens system for the wide and close 2L1S solutions.
For the prior probability distributions, we adopt the Galactic mass density and velocity distributions from the Galaxy model of \citet{Han+1995}, together with the stellar mass function of \citet{Sumi+2011}.
Because these priors are defined for single stars, we
rescale the event timescale and the Einstein radius to correspond
to those of the lens host.
The event timescale of the lens host, \( t_{\mathrm{E,H}} \), and the Einstein radius of the lens host, \( \theta_{\mathrm{E,H}} \), are expressed in terms of the mass ratio \( q \) as follows:


\begin{equation}
\label{eq:t_E,H}
    t_{\rm {E,H}}=\frac{t_{\rm E}}{\sqrt{1+q}},
\end{equation}

\begin{equation}\label{eq:theta_E,H}
    \theta_{\rm {E,H}}=\frac{\theta_{\rm E}}{\sqrt{1+q}}.
\end{equation}

We also estimate the apparent magnitudes of the lens host in the $V$, $I$, $H$, and $K$ bands, including the effects of extinction.
For this purpose, we adopt mass--luminosity relations for main-sequence stars based on \citet{Henry+1993} and \citet{Kroupa+1997}, together with the color--color relations of \citet{Kenyon+1995}.
For very low-mass stars and substellar objects ($M < 0.079\:M_\odot$), we adopt the 5~Gyr isochrone models of \citet{Baraffe+2003}.
The magnitudes at intermediate masses are obtained by interpolation.
The blending flux, $f_{\mathrm b}$, obtained from the light-curve modeling is adopted as the upper limit on the lens brightness.
When the lens magnitude inferred from the sampled lens mass is brighter than the blending magnitude in either the $V$ or $I$ band, converted using Equations~(\ref{eq:convert1}) and (\ref{eq:convert2}), we impose a Gaussian penalty based on the magnitude difference and its associated uncertainty.
Following \citet{Bennett+2015}, we estimate the extinction in front of the lens using the following equation:


\begin{equation}\label{eq:extinction}
    A_{i, \rm {L}}=\frac{1-\exp{[-D_{\rm L}/h_{\rm {dust}}}]}{1-\exp{[-D_{\rm S}/h_{\rm {dust}}}]}A_{i, \rm {S}},
\end{equation}

\noindent
where $i$ denotes the observed wavelength band, $A_{i,\mathrm{L}}$ and $A_{i,\mathrm{S}}$ are the total extinctions in the $i$ band toward the lens and the source, respectively, and $h_{\mathrm{dust}}$ is the dust scale length along the line of sight to the event.
The dust scale length is given by $h_{\mathrm{dust}} = (0.1\,\mathrm{kpc})/\sin |b|$, where $b$ is the Galactic latitude of the event.
We derive $A_{\mathrm{H}}$ and $A_{\mathrm{K}}$ from $A_{\mathrm{V}}$ using the wavelength dependence of the extinction law toward the Galactic center reported by \citet{Nishiyama+2008}.

\begin{figure*}
    \centering
    \hspace*{-20mm}
    \includegraphics[scale=0.7,angle=-90]{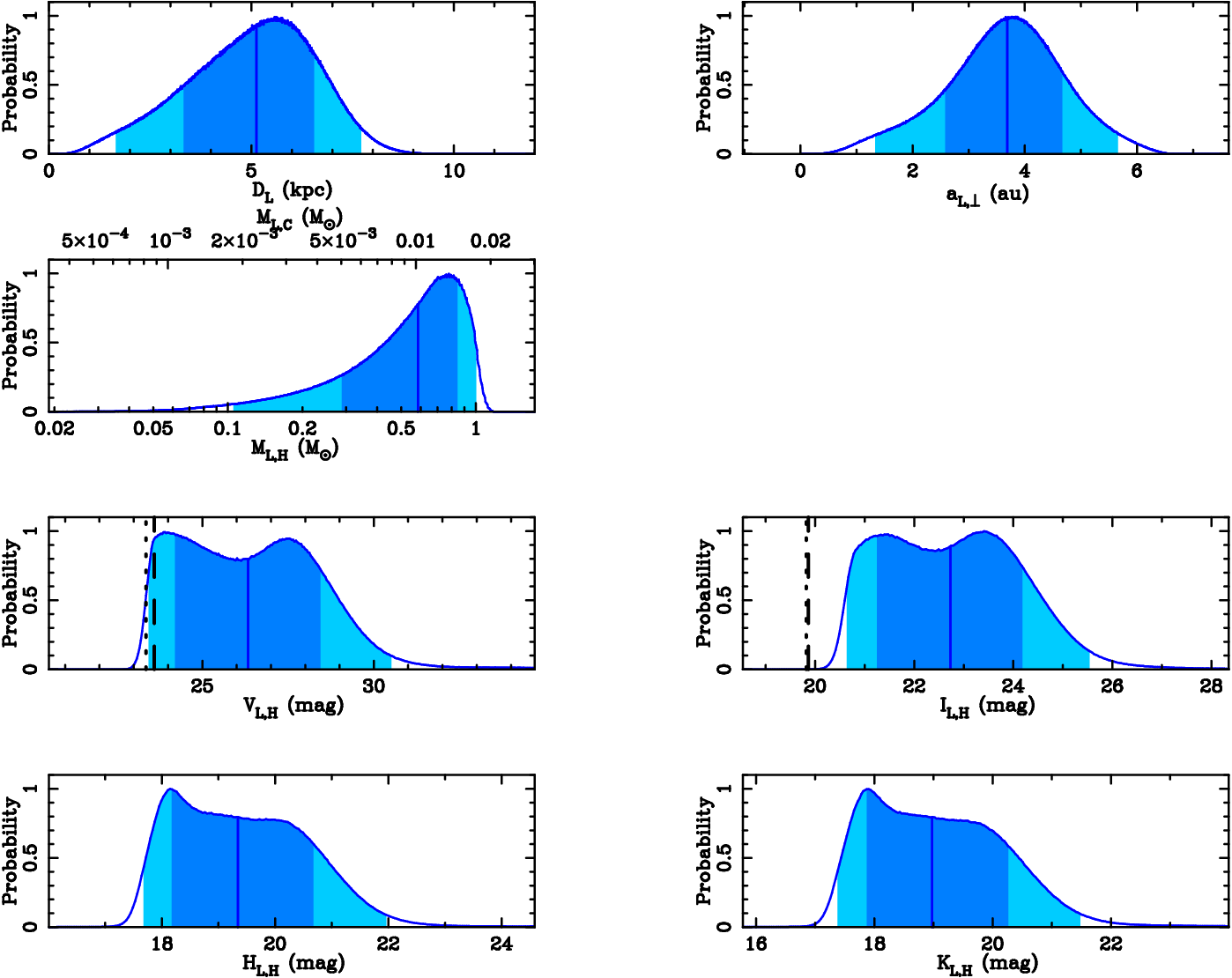}
    \vspace{15mm}
    \caption{
    Posterior probability distributions of the lens-system properties obtained from the Bayesian analysis for the wide 2L1S solution.
    In each panel, the dark blue and light blue shaded regions represent the $68.3\%$ and $95.4\%$ credible intervals, respectively, while the blue vertical line denotes the median value.
    For the apparent $V$- and $I$-band magnitudes, the dashed and dotted vertical lines indicate the blending magnitudes and their $1\sigma$ brighter limits, respectively, derived from the light-curve modeling.
    The blending magnitudes are adopted as upper limits on the lens brightness, and a Gaussian penalty is applied only to model-predicted lens magnitudes brighter than the corresponding blending magnitudes.
    \\
    {Alt text: Seven posterior probability distribution panels for the wide 2L1S solution. The horizontal axes show lens distance, projected host companion separation, lens host mass, and apparent V, I, H, and K band magnitudes of the lens host.}
    }
    \label{fig:Lens_2L1Swide}
\end{figure*}
\begin{figure*}
    \centering
    \hspace*{-20mm}
    \includegraphics[scale=0.7,angle=-90]{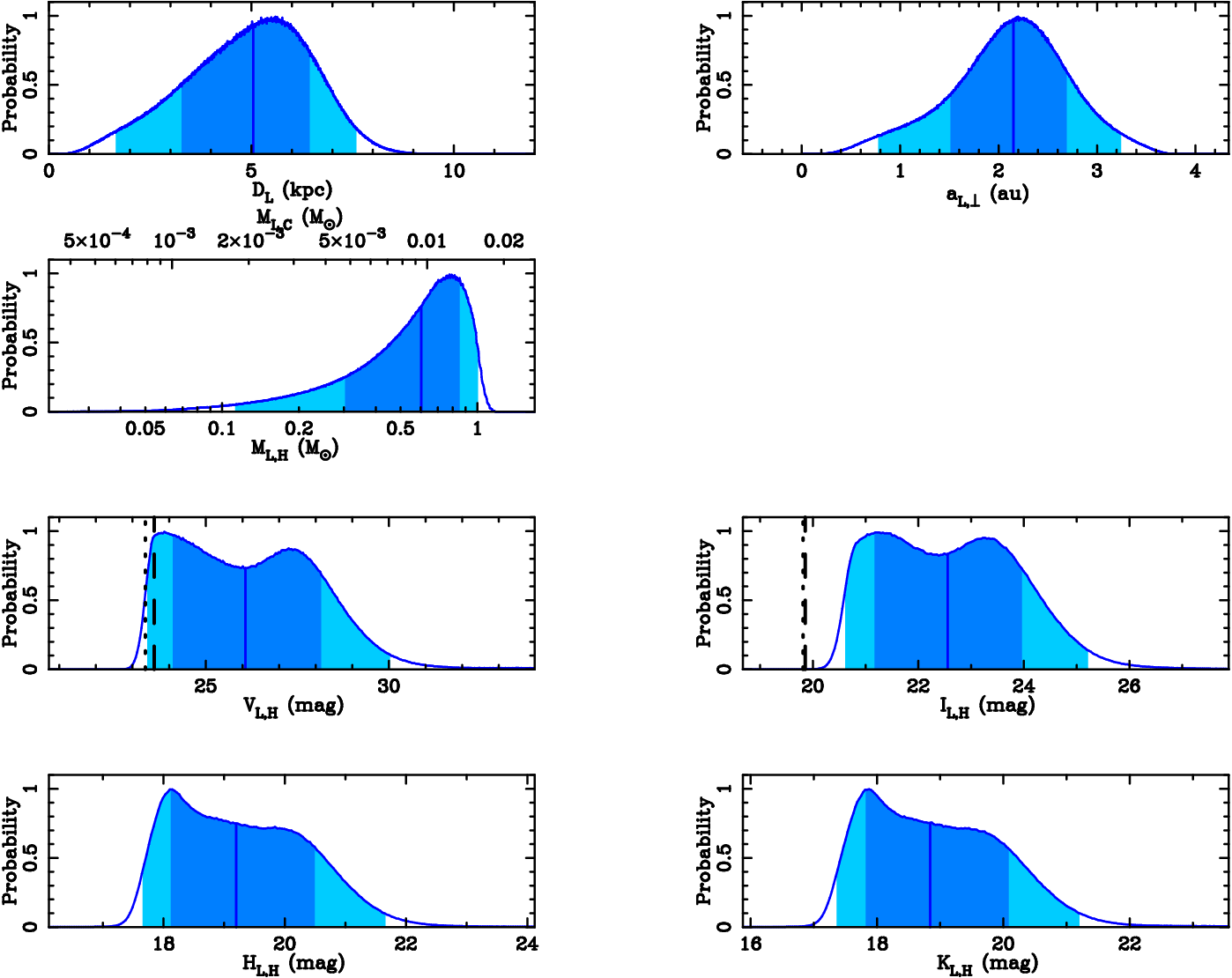}
    \vspace{15mm}
    \caption{
    Same as Figure~\ref{fig:Lens_2L1Swide}, but for the close 2L1S solution.\\
    {Alt text: Seven posterior probability distribution panels for the close 2L1S solution. The horizontal axes show lens distance, projected host companion separation, lens host mass, and apparent V, I, H, and K band magnitudes of the lens host.}
    }
    \label{fig:Lens_2L1Sclose}
\end{figure*}

Table~\ref{tab:Lens_Properties} summarizes the estimated physical properties of the lens system, including the distance to the lens, $D_{\mathrm{L}}$; the lens host mass, $M_{\mathrm{L,H}}$; the lens companion mass, $M_{\mathrm{L,C}}$; the projected orbital separation on the plane of the sky, $a_{\mathrm{L,\perp}}$; and the expected physical separation, $a_{\mathrm{L,exp}}$.
The table also lists the apparent magnitudes of the lens host in the four photometric bands, $V_{\mathrm{L,H}}$, $I_{\mathrm{L,H}}$, $H_{\mathrm{L,H}}$, and $K_{\mathrm{L,H}}$.
In addition, the blending magnitudes in the $V$ and $I$ bands, $V_{\mathrm{blend}}$ and $I_{\mathrm{blend}}$, which serve as upper limits on the total brightness of the lens system, are included.
The expected physical separation, $a_{\mathrm{L,exp}}$, is inferred from the projected separation, $a_{\mathrm{L,\perp}}$, through probabilistic deprojection, adopting the conditional probability distribution appropriate for random viewing orientations.
Figures~\ref{fig:Lens_2L1Swide} and \ref{fig:Lens_2L1Sclose} show the posterior probability distributions of the lens system properties for the wide and close 2L1S solutions, respectively.

The posterior distribution of the wide 2L1S solution indicates a $10.7_{-5.4}^{+4.8}$ $M_{\mathrm {Jup}}$ planet orbiting a $0.58_{-0.30}^{+0.26}$ $M_\odot$ host star with the expected orbital radius $a_{\mathrm {exp}} = 4.4_{-1.4}^{+2.1}$ au located $5.1_{-1.8}^{+1.4}$ kpc from Earth.
Meanwhile, the posterior distribution of the close 2L1S solution indicates a $9.9_{-4.9}^{+4.1}$ $M_{\mathrm {Jup}}$ planet orbiting a $0.60_{-0.30}^{+0.25}$ $M_\odot$ host star with the expected orbital radius $2.6_{-0.8}^{+1.2}$ au located $5.0_{-1.8}^{+1.4}$  kpc from Earth.
The results of both models are similar in all parameters except for the orbital radius, and are consistent in that the planet is a giant planet of about $10$ Jupiter masses orbiting a host star of about $0.6$ solar masses at a distance of about $5$ kpc from Earth.
Considering the expected orbital radius, the companion is likely located beyond the snow line of the host star, given by $\sim 2.7\:\mathrm {au}\:(M/M_\odot) \sim 1.6 \:\mathrm {au}$.

\begin{table}
  \tbl{Lens Properties of the 2L1S Models
}{
  \begin{tabular}{lccc}
      \hline
      Parameters & Units & wide 2L1S & close 2L1S   \\[3pt] 
      \hline
      $D_{\mathrm L}$ & $\mathrm {kpc}$ & $5.13_{-1.80}^{+1.42}$ & $5.04_{-1.76}^{+1.40}$  \\[3pt]
      $M_{\mathrm {L,H}}$ & $M_\odot$ & $0.58_{-0.30}^{+0.26}$ & $0.60_{-0.30}^{+0.25}$ \\[3pt]
      $M_{\mathrm {L,C}}$ & $M_\mathrm{Jup}$ & $10.67_{-5.41}^{+4.75}$ & $9.93_{-4.93}^{+4.12}$ \\[3pt]
      $a_{\mathrm {L,\perp}}$ & $\mathrm {au}$ & $3.69_{-1.11}^{+0.98}$ & $2.15_{-0.64}^{+0.54}$ \\[3pt]
      $a_{\mathrm {L, exp}}$ & $\mathrm {au}$ & $4.44_{-1.41}^{+2.13}$ & $2.58_{-0.81}^{+1.21}$ \\[3pt]
      $V_{\mathrm {L,H}}$ & $\mathrm {mag.}$ & $26.34_{-2.13}^{+2.11}$ & $26.08_{-1.98}^{+2.07}$ \\[3pt]
      $I_{\mathrm {L,H}}$ & $\mathrm {mag.}$ & $22.73_{-1.48}^{+1.45}$ & $22.55_{-1.38}^{+1.41}$ \\[3pt]
      $H_{\mathrm {L,H}}$ & $\mathrm {mag.}$ & $19.34_{-1.17}^{+1.33}$ & $19.20_{-1.08}^{+1.29}$ \\[3pt]
      $K_{\mathrm {L,H}}$ & $\mathrm {mag.}$ & $18.98_{-1.10}^{+1.29}$ & $18.84_{-1.02}^{+1.24}$ \\[3pt]
      $V_{\mathrm {Blend}}$ & $\mathrm {mag.}$ & $23.60\pm0.24$ & $23.59\pm0.24$ \\[3pt]
      $I_{\mathrm {Blend}}$ & $\mathrm {mag.}$ & $19.86\pm0.04$ & $19.86\pm0.04$ \\[3pt]
      \hline
    \end{tabular}}\label{tab:Lens_Properties}
\end{table}

\section{Discussion and Conclusion}
\label{sec:Discussion_and_Conclusion}

We report an analysis of the gravitational microlensing event MOA-2020-BLG-108.
By modeling the observed light curve, we found that the lens is a planetary system with a mass ratio of $q\sim 0.02$.
We found two degenerate 2L1S solutions corresponding to the wide and close configurations.
No significant higher-order effects were detected, whereas the finite-source parameter was constrained in the final models.
We therefore adopted the wide and close 2L1S solutions as the final models.
We then conducted a Bayesian analysis to estimate the physical parameters of the lens system and found that both the wide and close 2L1S solutions correspond to a giant planet with a mass of approximately $10\:M_{\rm Jup}$ orbiting a host star of about $0.6\:M_\odot$ at a distance of about $5\,\mathrm{kpc}$ from Earth.
Moreover, the companion is likely orbiting outside the snow line.

Galactic kinematic models show that disk–bulge microlensing events toward the Galactic bulge typically have larger relative proper motions ($\sim 5$--$10~\mathrm{mas\,yr^{-1}}$) than bulge–bulge self-lensing events ($\sim 2$--$5~\mathrm{mas\,yr^{-1}}$) \citep[e.g.,][]{Kiraga+1994,Han+2003, Dominik+2010, Koshimoto+2021a}.
In this analysis, the relative proper motion is $ \sim 7.5 \: \rm{mas} \:\rm{yr}^{-1}$ for the wide 2L1S solution and $ \sim 7.0 \: \rm{mas} \:\rm{yr}^{-1}$ for the close 2L1S solution, suggesting that MOA-2020-BLG-108 is likely a disk–bulge event.

In microlensing analyses, it is well known that the close–wide degeneracy, i.e., the $s \leftrightarrow 1/s$ degeneracy, can arise \citep[e.g.,][]{Dominik1999, Griest+1998, Chung+2005}.
The physical parameters inferred from the two degenerate solutions are largely consistent, making it difficult to distinguish between them even with future observations.
Statistical studies based only on events with uniquely determined solutions may be subject to selection effects such as publication bias.
Therefore, even in the presence of degenerate solutions, it is ideal to include them in statistical analyses by applying weights based on their likelihood distributions.
Because MOA-2020-BLG-108Lb is a potential sample for future statistical studies, properly reporting both degenerate solutions is important for constructing less biased population distributions of microlensing planets.

In general, microlensing light curves constrain the mass ratio more precisely than the absolute mass of the lens.
According to the IAU working definition of an exoplanet, a planet is required not only to have a mass below the deuterium-burning limit, but also to satisfy the companion-to-host mass-ratio criterion of $q < 2/(25+\sqrt{621})\simeq 0.04$ \citep{Lecavelier_des_Etangs+2022}.
For this event, the best-fit mass ratio and its $1\sigma$ range lie
below the IAU mass-ratio criterion.
On the other hand, while the median companion mass inferred from the Bayesian analysis is $10\:M_{\rm Jup}$, the upper bound of the $68.3\%$ credible interval exceeds $13\:M_{\rm Jup}$, the conventional deuterium-burning limit often used to distinguish planets from brown dwarfs.
Therefore, the classification of MOA-2020-BLG-108Lb as either a brown dwarf or a planet remains ambiguous.
The formation pathways of objects in this mass regime may be explained by several scenarios, including the core accretion scenario \citep[e.g.,][]{Mizuno1980, Pollack+1996}, the disk instability scenario—in which a protoplanetary disk fragments due to its own gravitational instability and directly forms planets or companions \citep[e.g.,][]{Boss1997, Mayer+2002}—and the gravitational collapse scenario, in which objects form via the fragmentation of molecular clouds \citep[e.g.,][]{Padoan+2002, Hennebelle+2008}.

As discussed in Section~\ref{sec:Introduction}, several statistical studies of companion mass ratios \citep{Zhang2025, Duchene+2023} have reported evidence for a mass-ratio desert in the range $0.02 \lesssim q \lesssim 0.05$, and suggested that the mass ratio, rather than the absolute mass, may be a more relevant parameter for understanding formation mechanisms.
The estimated mass ratio of this event is $q\sim0.02$, which lies near the lower boundary of the mass-ratio desert, or equivalently, near the upper boundary of the planetary population.
We note, however, that the evolutionary state of the lens host is not directly determined in the present analysis. 
If the lens host were a white dwarf, its progenitor would have been more massive than the present lens host, and the companion-to-host mass ratio at the time of formation would have been smaller.
The white dwarf mass function in the Galactic disk is known to exhibit a sharp peak near $\sim 0.6\:M_\odot$ \citep[e.g.,][]{Liebert+2005, Kepler+2007}, and the median host mass inferred from our Bayesian analysis is also approximately $0.6\:M_\odot$.
Using the initial-final mass relation for white dwarfs derived by \citet{Cunningham+2024}, a white dwarf of $0.6\:M_\odot$ corresponds to a progenitor mass of approximately $1.5\:M_\odot$.
In this case, the mass ratio during the main-sequence phase would have been $\sim 6 \times 10^{-3}$.
Future observations capable of detecting the lens flux could test this scenario.
If the detected flux is consistent with the properties expected for a main-sequence host, as summarized in Table~\ref{tab:Lens_Properties}, the white dwarf host scenario would be disfavored.
In any case, accumulating a larger sample of companions in this mass-ratio regime is crucial for advancing our understanding of the formation of giant planets and brown dwarfs.

The Nancy Grace Roman Space Telescope is expected to detect approximately 1400 planetary microlensing events, including many wide-orbit exoplanets with masses ranging from giant planets to below the Earth mass, as well as a large number of binary-lens events. 
Thanks to its excellent photometric precision, high angular resolution, and 12-minute observing cadence, Roman will often allow the lens masses to be constrained through a combination of light-curve and imaging observables, in addition to precisely measuring the mass ratios \citep[e.g.,][]{Spergel+2015,Bachelet+2024,Terry+2026}.
Such a large sample, for which mass ratios and, in many cases, lens masses can be constrained, will enable the mass-ratio distribution near the planet--brown-dwarf boundary to be measured more precisely.

MOA-2020-BLG-108 also lies within the currently planned footprint of the Roman Galactic Plane Survey \citep{GPS+2025} and is therefore expected to be imaged at high angular resolution by Roman.
Given the measured lens--source relative proper motion of $\sim 7\ {\rm mas\ yr^{-1}}$, the lens and source are expected to be separated by several tens of milliarcseconds by the time of the Roman observations.
If the lens host is a main-sequence star, the lens and source are expected to have comparable brightnesses, as indicated by Tables~\ref{tab:Source_Properties} and \ref{tab:Lens_Properties}.
The combination of this separation and the modest lens--source flux contrast may allow image elongation to be detected in Roman imaging.
Such a measurement could constrain the lens flux and yield an independent estimate of the lens--source relative proper motion, thereby improving the constraints on the masses of the lens host and its companion \citep[e.g.,][]{Bhattacharya+2018}.

\section*{Funding}
Y.K.S. was supported by JSPS KAKENHI Grant Number JP24K17092.
The MOA project is supported by JSPS KAKENHI Grant Numbers JP23340064, JP24253004, JP26247023, JP15H00781, JP16H06287, JP17H02871, JP19KK0082, JP22H00153, and JP23KK0060, and by the JSPS Core-to-Core Program (Grant Numbers JPJSCCA20210003 and JPJSCCA20260002).

\bibliographystyle{apj}
\bibliography{reference}

\end{document}